\documentclass[a4paper,preprints,article,accept,moreauthors,pdftex]{Definitions/mdpi} 

\firstpage{1} 
\pubvolume{1}
\issuenum{1}
\articlenumber{0}
\pubyear{2021}
\copyrightyear{2021}
\datereceived{12 April 2021} 
\dateaccepted{8 May 2021} 
\datepublished{} 
\hreflink{$\;$} 
\pdfoutput=1

\newcommand{\dd}{{\rm d}}

\newcommand{\Lag}{\mathcal{L}}

\newcommand{\mS}{\mathcal{S}}

\newcommand{\D}{\mathcal{D}}
\newcommand{\mR}{\mathcal{R}}
\newcommand{\mU}{\mathcal{U}}

\newcommand{\be}{\begin{equation}}
\newcommand{\ee}{\end{equation}}
\newcommand{\bea}{\begin{eqnarray}}
\newcommand{\eea}{\end{eqnarray}}

\newcommand{\mD}{{\mathcal D}}

\newcommand{\mH}{{\mathcal H}}

\newcommand{\mpl}{M_{\rm Pl}}
\newcommand{\TT}{\mathbb{T}}
\newcommand{\GG}{\mathbb{G}}
\newcommand{\QQ}{\mathbb{Q}}

\newcommand{\GTGR}{\mathring{\mathbb{G}}}

\newcommand{\Qb}{\bar{Q}}
\newcommand{\phib}{\bar{\phi}}
\newcommand{\psib}{\bar{\psi}}

\newcommand{\UE}{\mathcal{U}_{\rm E}}

\newcommand{\GL}{GL(4,\mathbb{R})}

\renewcommand{\so}{\mathfrak{so}(3,1)}

\newcommand{\LambdaE}{\tilde{\Lambda}}
\newcommand{\TE}{\tilde{T}}

\Title{Accidental gauge symmetries of Minkowski spacetime in Teleparallel theories.}

\Author{Jose Beltr\'an Jim\'enez$^{1}$ and Tomi S. Koivisto$^{2,3}$}

\address{
$^{1}$ \quad Departamento de F\'isica Fundamental and IUFFyM, Universidad de Salamanca, E-37008 Salamanca, Spain; jose.beltran@usal.es\\
$^{2}$ \quad Laboratory of Theoretical Physics, Institute of Physics, University of Tartu, W. Ostwaldi 1, 50411 Tartu, Estonia; tomi.koivisto@ut.ee \\
$^{3}$ \quad National Institute of Chemical Physics and Biophysics, R{\"a}vala pst. 10, 10143 Tallinn, Estonia; tomi.koivisto@kbfi.ee}

\abstract{In this paper, we provide a general framework for the construction of the Einstein frame within non-linear extensions of the teleparallel equivalents of General Relativity. These include the metric teleparallel and the symmetric teleparallel, but also the general teleparallel theories. We write the actions in a form where we separate the Einstein--Hilbert term, the conformal mode due to the non-linear nature of the theories (which is analogous to the extra degree of freedom in $f(R)$ theories), and the sector that manifestly shows the dynamics arising from the breaking of local symmetries. This frame is then used to study the theories around the Minkowski background, and we show how all the non-linear extensions share the same quadratic action around Minkowski. As a matter of fact, we find that the gauge symmetries that are lost by going to the non-linear generalisations of the teleparallel General Relativity equivalents arise as accidental symmetries in the linear theory around Minkowski. Remarkably, we also find that the conformal mode can be absorbed into a Weyl rescaling of the metric at this order and, consequently, it disappears from the linear spectrum so only the usual massless spin 2 perturbation propagates. These findings unify in a common framework the known fact that no additional modes propagate on Minkowski backgrounds, and we can trace it back to the existence of accidental gauge symmetries of such a background.}

\begin{document}
\end{paracol}
\section{Introduction}

Besides the standard description of General Relativity (GR) in terms of the
metric and its Levi-Civita connection, the~theory has alternative formulations
in terms of flat connections~\cite{BeltranJimenez:2019tjy}. The~reformulation using a flat and
metric-compatible connection, ``Einstein's 2nd GR'', is known as the
Teleparallel Equivalent of GR (TEGR) \cite{Aldrovandi:2013wha}. The~reformulation in terms of a flat
and symmetric connection, known as the Symmetric Teleparallel Equivalent of
GR (STEGR) \cite{Nester:1998mp}, has been established as the minimal covariantisation of Einstein
1st GR, ``the $\Gamma\Gamma$ formulation'', and~as such the unique realisation
of gravity as the gauge theory of translations~\cite{BeltranJimenez:2017tkd}. The~possible fundamental
motivation for teleparallelism could apparently be very simple: the Planck
mass is the mass of the gravitational connection~\cite{Koivisto:2019jra}. This would also explain the
otherwise anomalous dimension of the GR action that renders it
non-renormalisable, and~provides a completely new approach towards a UV completion of~gravity.

In the light of these developments, the~current interest in teleparallel
gravity, which is also reflected in the several contributions to this Special
Issue, is well justified. However, most of these studies, including the
one at hand, focus on non-linear modifications of the teleparallel equivalents
of GR (which, from~the perspective of teleparallelism as the low-energy
manifestation of the ultra-massive spacetime connection, could perhaps be
interpreted as non-linear extensions of the quadratic Proca-like term) which have
been mainly motivated by their potential use as models of cosmological inflation and dark energy (and even dark matter~\cite{Milgrom:2019rtd,DAmbrosio:2020nev}). The~non-linear extension of TEGR, the~$f(\TT)$ theory~\cite{Bengochea:2008gz}, has been considered in, e.g.,~\cite{Li:2010cg,Li:2011rn,Ferraro:2018axk,Ferraro:2018tpu,Jarv:2019ctf,Hohmann:2019nat,Hashim:2020sez,Golovnev:2020nln,Jimenez:2020ofm,Blagojevic:2020dyq}, and~the non-linear extension of STEGR, the~$f(\QQ)$ \cite{BeltranJimenez:2017tkd} and related theories have been considered in, e.g.,
~\cite{Jimenez:2019ovq,Lazkoz:2019sjl,Mandal:2020buf,Ayuso:2020dcu,Xu:2020yeg,Bajardi:2020fxh,Flathmann:2020zyj,Frusciante:2021sio,Yang:2021fjy,Khyllep:2021pcu}. 
The general teleparallel equivalent of GR, not subject to either the metric or the symmetric condition, was introduced quite recently~\cite{Jimenez:2019ghw}, and~only in this paper do we take some first steps towards understanding the properties of the non-linearly extended $f(\GG)$ theory. The~aim of the note is to consider the three classes of theories from the perspective of the Einstein frame, and~in particular to clarify their common feature, which is the evanescence of the extra degrees of freedom from the linear spectrum on a Minkowski~background.   

The physical content of these gravity theories in arbitrary spacetimes could be properly resolved in the non-perturbative framework of Hamiltonian analysis. This has indeed been pursued by several authors, though~exclusively in the context of metric teleparallel gravity~\cite{Li:2011rn,Ferraro:2018tpu,Blagojevic:2020dyq,Blixt:2019mkt}.
The discussion is converging towards the conclusion that there may exist more
than one extra degree of freedom, and~that this can depend on the location in
the phase space in such a way that the possible physical interpretation
remains a thorny issue~\cite{Blixt:2020ekl}. One may be inclined to simply
discard the modified metric teleparallel models as unphysical due to their
violation of fundamental symmetry~\cite{Kopczynski_1982,Li:2010cg}, but~from
a theoretical perspective, a more thorough understanding of the problem
and its repercussions is desirable, and, arguably prompted by the
neat performance of the models in some phenomenological aspects (see, e.g.,~\cite{Hashim:2020sez} of the many references listed above). As~the necessary first step to uncovering the nature of the degrees of freedom in the non-linearly extended theories, we scrutinise their Minkowski space limit, but~we are also able to make some generic conclusions based on the Einstein frame formulations developed in this~note.

The general teleparallel equivalent of GR is reviewed in Section~\ref{sec:generalT}, where it is shown that the TEGR and the STEGR are recovered as the gauge-fixed versions of the general theory. In~Section~\ref{sec:framing}, we construct the scalar--tensor formulation and the 
corresponding Einstein frame of $f(\QQ)$ gravity by making the suitable conformal transformation~\cite{Jarv:2018bgs,Iosifidis:2018zwo,Gakis:2019rdd}, and~exploit this formulation to gain insights into the general structure, the~cosmology and, finally, the~Minkowski space limit of the theory. In~Section~\ref{Sec:MetricT}, we construct the analogous Einstein frame formulation of the metric teleparallel $f(\TT)$ theory~\cite{Wright:2016ayu,Ferraro:2018axk,Raatikainen:2019qey}, and~in
\mbox{Section~\ref{sec:general}}, the same for the general teleparallel $f(\GG)$ theory. The~latter provides a unified framework to consider both the previously studied classes of theories. In~Section~\ref{sec:Minkowski}, we consider the more specific problem of uniqueness of the Minkowski space. Section~\ref{sec:discussion} summarises the findings in this~paper.  
\\

{\it{Notation}:} We will use the mostly plus signature of the metric. The~covariant derivative of the general connection $\Gamma$ will be $\nabla$, while the Levi-Civita covariant derivative will be $\mD$. The~curvatures will be denoted as $R$ and $\mR$ for the general and Levi-Civita connections, respectively. The~non-metricity is defined as $Q_{\alpha\mu\nu}=\nabla_\alpha g_{\mu\nu}$ with its two independent traces $Q_\mu=g^{\alpha\beta} Q_{\mu\alpha\beta}$ and $\Qb_\alpha=g^{\mu\beta}Q_{\mu\alpha\beta}$. The~torsion is defined as $T^\alpha{}_{\mu\nu}=2\Gamma^\alpha{}_{[\mu\nu]}$ and its trace as $T_\mu=T^\alpha{}_{\mu\alpha}$.

\section{The General Teleparallel~Gravity}
\label{sec:generalT}

We will start by establishing the general framework to be used for the teleparallel theories that we will consider in this work. The~underlying idea to obtain teleparallel equivalents of GR resides in the post-Riemannian expansion of the Ricci scalar for a general connection $\Gamma$ given by
\be
R=\mR-\GTGR+\mD_{\mu}\left(Q^{\mu}-\Qb^{\mu}+2 T^{\mu}\right)
\label{eq:RtoROmega2}
\ee
where we have defined the general teleparallel quadratic scalar~\cite{Jimenez:2019ghw} (see also~\cite{Iosifidis:2018zwo,Iosifidis:2019dua})
\begin{align}
-\GTGR=&\frac{1}{4} T_{\mu \nu \rho} T^{\mu \nu \rho}+\frac{1}{2} T_{\mu \nu \rho} T^{\mu \rho \nu}-T_{\mu} T^{\mu}+Q_{\mu \nu \rho} T^{\rho\mu \nu}-Q_{\mu} T^{\mu}+\bar{Q}_{\mu} T^{\mu}\nonumber\\
&+\frac{1}{4} Q_{\mu \nu \rho} Q^{\mu \nu \rho}-\frac{1}{2} Q_{\mu \nu \rho} Q^{\nu \mu \rho}-\frac{1}{4} Q_{\mu} Q^{\mu}+\frac{1}{2} Q_{\mu} \bar{Q}^{\mu}.
\end{align}

The decomposition \eqref{eq:RtoROmega2} clearly shows how the Ricci scalar of the Levi-Civita connection differs from the teleparallel scalar $\GTGR$ by a divergence term for a flat connection with $R=0$. This property roots the construction of teleparallel equivalents of GR because the Einstein--Hilbert action can then be equivalently expressed in terms of $\GTGR$ up to a boundary term that is irrelevant for the classical equations of motion. More explicitly, the~two actions
\begin{align}
\mS_{\rm EH}[g]=\frac12\mpl^2\int\dd^4x\sqrt{-g}\mR(g)\quad{\rm and}\quad \mS_{{\rm GR}_\parallel}[g,\Gamma]=\frac12\mpl^2\int\dd^4x\sqrt{-g}\;\GTGR
\end{align}
characterise the same dynamical system. As~we know, GR describes two propagating degrees of freedom (DOFs) corresponding to the two polarisations of the gravitational waves encoded into the spacetime metric $g_{\mu\nu}$, and~incorporates the gauge symmetry provided by diffeomorphisms. The~teleparallel description, however, contains the metric and the connection DOFs, so we need additional ingredients if they are to describe the same two DOFs of GR. We still have diffeomorphisms invariance since the action $ \mS_{{\rm GR}_\parallel}$ is constructed as a scalar. However, the~teleparallel scalar $\GTGR$ is special among the general class of quadratic teleparallel actions because it enjoys an additional $\GL$ local symmetry for flat geometries~\cite{Jimenez:2019ghw}. To~understand this statement more clearly, let us first notice that the flatness constraint imposes the connection to be purely inertial, i.e.,~it is given by a pure gauge mode that we can express as
\be
\Gamma^\alpha{}_{\mu\beta}=(\Lambda^{-1})^\alpha{}_\rho\partial_\mu\Lambda^\rho{}_\beta
\label{eq:Telecon}
\ee
with $\Lambda^\alpha{}_\beta$ an arbitrary element of $\GL$. The~exceptional property of the scalar $\GTGR$ is that, when evaluated on the connection \eqref{eq:Telecon}, it does not contribute to the dynamics of the action described by $\mS_{{\rm GR}_\parallel}$. The~equivalence to GR can be understood from the fact that $\Lambda^\alpha{}_\mu$ only enters as a total derivative in $\mS_{{\rm GR}_\parallel}[g,\Lambda]$, so the derived equations for the connection are trivial, while the metric field equations are oblivious to $\Lambda^\alpha{}_\mu$ and are in fact the same as those derived from $\mS_{\rm EH}[g]$. We refer to~\cite{Jimenez:2019ghw} for a more detailed derivation of these statements. The~relevant property for our study in this note is that the disappearance of the inertial connection can be interpreted as the presence of an additional $\GL$ local symmetry. In~any case, having at our disposal the general local $\GL$, we can make different gauge choices, and two of them stand out in the literature for their clear geometrical interpretation: 

\begin{itemize}
\item{ \it The metric teleparallel gauge}. This gauge is defined by further imposing that the non-metricity of the connection vanishes $Q_{\alpha\mu\nu}=0$ and the torsion is the only non-trivial object associated to the connection. The~non-metricity constraint imposes a relation between $\Lambda$ and the metric that can be solved as $g_{\mu\nu}=\Lambda^\alpha{}_\mu \Lambda^\beta{}_\nu \eta_{\alpha\beta}$, where we have assumed that the metric reduces to Minkowski for the identity in $\GL$ (see~e.g.,~\cite{Jimenez:2019tkx,Jimenez:2019ghw} for further discussions on this point). The~teleparallel action in this gauge defines the TEGR theory, and the fundamental scalar is given by $\TT=\GTGR(Q^\alpha{}_{\mu\nu}=0)$.

\item{\it The symmetric teleparallel gauge}. In~this gauge, the~torsion is trivialised so we have the condition $T^\alpha{}_{\mu\nu}=0$. This condition forces the inertial connection to be generated by an element of $\GL$ of the form $\Lambda^{\alpha}{}_\mu=\partial_\mu \xi^\alpha$ for some arbitrary functions $\xi^\alpha$. This is exactly what corresponds to a transformation of the connection under a diffeomorphism so that one can completely remove the connection by using appropriate coordinates. This system of coordinates is $\xi^\alpha=x^\alpha$ (modulo a global affine transformation) that is called the unitary or coincident gauge. The~theory in this gauge defines the STEGR or Coincident GR (alluding to the gauge choice) and is described by the non-metricity scalar $\QQ=\GTGR(T^\alpha{}_{\mu\nu}=0)$.
\end{itemize}

Non-linear extensions based on the above two gauge choices~\cite{Bengochea:2008gz,BeltranJimenez:2017tkd} have been considered in the literature at some length~\cite{Li:2010cg,Li:2011rn,Ferraro:2018axk,Ferraro:2018tpu,Jarv:2019ctf,Hohmann:2019nat,Hashim:2020sez,Golovnev:2020nln,Jimenez:2020ofm,Blagojevic:2020dyq,Jimenez:2019ovq,Lazkoz:2019sjl,Mandal:2020buf,Ayuso:2020dcu,Xu:2020yeg,Bajardi:2020fxh,Flathmann:2020zyj,Frusciante:2021sio,Yang:2021fjy,Khyllep:2021pcu}. Regarding these generalisations, it is important to notice that the presence of the different boundary terms is what makes the non-linear extensions based on different gauge choices give rise to different theories. In~the next sections, we will discuss how they can be expressed in appropriate frames where the resemblance to GR is apparent and also the deviations. In~particular, the~conformal mode can be isolated, and we will see how it couples to the boundary terms that break the corresponding local~symmetries.

\section{Framing Symmetric~Teleparallelisms}
\label{sec:framing}
As discussed above, the~teleparallel scalar $\GTGR$ differs from the Ricci scalar of the Levi-Civita connection by a divergence. This is innocuous for the dynamics of theories that are linear in these scalars, but~it will have important effects on non-linear extensions. These extensions have been explored at length in the literature and different frames have been constructed to unveil some of their properties. For~$f(\mR)$ theories, it is well-known that an Einstein frame exists where the additional scalar DOF can be made explicit as a conformal mode. The~same construction lacks in the teleparallel non-linear extensions. Here, we will construct a frame where the differences with respect to GR are most apparent so that we can clearly isolate the novel effects, mainly the new DOFs and the sector responsible for the breaking of the symmetries. Let us start by considering the $f(\QQ)$ theories described by:
\be
\mS=\frac12 \mpl^2\int\dd^4x\sqrt{-g} f(\QQ)
\ee
 with $f$ some function of the non-metricity scalar. We will exploit the relation (\ref{eq:RtoROmega2}) reducing now to
\be
\QQ=\mR(g)+\D_\alpha J^\alpha
\ee
with $J^\mu\equiv Q^\mu-\Qb^\mu$, to~rewrite the theory in a form closer to the more common $f(\mR)$ theories and go as close as possible to the GR formulation so that we can clearly isolate the Diff-violating sector of the theory. We can perform a Legendre transformation to write the action as
\be
\mS=\frac12 \mpl^2\int\dd^4x\sqrt{-g}\left[f(\chi)+\varphi\Big(\mR+\D_\mu J^\mu-\chi\Big)\right].
\ee

We have introduced two auxiliary fields, $\varphi$ and $\chi$, that can be integrated out to recover the original form of the action. To~arrive at our desired frame, we can integrate out only the field $\chi$ from its equation of motion $f'(\chi)=\varphi$ so the action can be expressed as
\be
\mS=\frac12 \mpl^2\int\dd^4x\sqrt{-g}\Big[\varphi\mR-\mU(\varphi)-\partial_\alpha\varphi J^\alpha\Big].
\label{eq:BransDickeF(Q)}
\ee
with 
\be
\mU=\Big[\varphi\chi-f\Big]_{\chi=\chi(\varphi)}
\label{SCG0}
\ee
and we have also integrated by parts in the last term. This frame resembles the Jordan--Brans--Dicke frame of classical scalar--tensor theories, except for the coupling of $\varphi$ to the current $J^\alpha$ that is responsible for the crucial differences. 
To go to the Einstein frame, we now perform a conformal transformation $g_{\mu\nu}=\frac{1}{\varphi} q_{\mu\nu}$ to recover the standard Einstein--Hilbert term. After~some arrangements and introducing the field redefinition $\varphi\equiv e^{2\phi}$, the~action can be written as
\be
\mS=\frac12 \mpl^2\int\dd^4x\sqrt{-q}\left[\mR(q)-6(\partial\phi)^2-\tilde{\mU}(\phi)-2e^{-2\phi}\partial_\alpha\phi J^\alpha\right]
\ee
 where we have defined $\tilde{\mU}=e^{-4\phi}\mU$. We already have the Einstein--Hilbert sector, but~to achieve the final form of the Einstein frame, we still need to perform the conformal transformation for the current $J^\mu$ to express it in terms of the transformed metric $q_{\mu\nu}$. A~simple computations yields
\be
J^{\alpha}=\Big(g^{\alpha\beta}g^{\mu\nu}-g^{\alpha\mu}g^{\beta\nu}\Big)\nabla_\beta g_{\mu\nu}=e^{2\phi}\left[\Big(q^{\alpha\beta}q^{\mu\nu}-q^{\alpha\mu}q^{\beta\nu}\Big)\nabla_\beta q_{\mu\nu}-6q^{\alpha\beta}\partial_\beta\phi\right].
\ee

With this expression, we can finally write the action in the following apparent form
\be
\mS=\frac12 \mpl^2\int\dd^4x\sqrt{-q}\left[\mR(q)+6(\partial\phi)^2-\tilde{\mU}(\phi)-2\Big(q^{\alpha\beta}q^{\mu\nu}-q^{\alpha\mu}q^{\beta\nu}\Big)\partial_\alpha\phi\nabla_\beta q_{\mu\nu}\right]
\ee
that is the Einstein frame representation of the $f(\QQ)$ theory. To date, we have not fixed any gauge, so we are still free to choose a suitable one. As~usual, a~convenient choice is the unitary or coincident gauge where the connection trivialises so $\nabla\rightarrow \partial$. We should notice that we did not have to transform the connection since it is an independent field that we can trivialise. Of~course, we could have assigned some transformation property for the $\xi$'s, but~our results do not depend on this and, at~this stage, it would seem like an unnecessary complication (see nonetheless the relevant discussion in Section~\ref{sec:generalT}). Therefore, we can write
\begingroup\makeatletter\def\f@size{9}\check@mathfonts
\def\maketag@@@#1{\hbox{\m@th\normalsize\normalfont#1}}%

\be
\mS_{\rm CG}=\int\dd^4x\sqrt{-q}\left[\frac12 \mpl^2\mR(q)+3(\partial\phi)^2-\frac12 \mpl^2\tilde{\mU}(\phi)-\mpl\Big(q^{\alpha\beta}q^{\mu\nu}-q^{\alpha\mu}q^{\beta\nu}\Big)\partial_\alpha\phi\partial_\beta q_{\mu\nu}\right],
\label{SCG}
\ee
\endgroup
where we have also restored the natural dimension of the scalar field $\phi\rightarrow \phi/\mpl$. This formulation of the theory shows in a very transparent manner the solutions with a constant conformal mode coincides with those of GR up to a shift in the cosmological constant originating from the potential $\tilde{\mU}$. Thus, differences with respect to GR will only appear for Lorentz breaking configurations of the conformal~mode.

The action can be written yet in an alternative, possibly more useful for some developments, manner by using the relations
\be
q^{\alpha\beta} q^{\mu\nu}\partial_\beta q_{\mu\nu}=\partial^\alpha\log \vert q\vert\quad\text{and}\quad q^{\alpha\mu} q^{\beta\nu}\partial_\beta q_{\mu\nu}=-\partial_\beta q^{\beta\alpha}
\ee
with $q=\det q_{\mu\nu}$. We can then express the action in the following equivalent form:
\begingroup\makeatletter\def\f@size{9}\check@mathfonts
\def\maketag@@@#1{\hbox{\m@th\normalsize\normalfont#1}}%
\be
\mS_{\rm CG}=\int\dd^4x\sqrt{-q}\left[\frac12 \mpl^2\mR(q)+3(\partial\phi)^2-\frac12 \mpl^2\tilde{\mU}(\phi)-\mpl\Big(\partial^\alpha\log \vert q\vert+\partial_\beta q^{\beta\alpha}\Big)\partial_\alpha\phi\right].
\ee
\endgroup

It is interesting to note that the first term of the Diff-breaking sector $\partial^\alpha\log q\partial_\alpha\phi$ respects volume-preserving-Diffs. In~this respect, it would be interesting to explore potential relations with unimodular gravities and its~deformations.

As promised, we have rewritten the theory in a form that closely resembles the usual formulation of GR, and we can see that the effect of considering a non-linear function of the non-metricity scalar is twofold: the appearance of the dynamical conformal mode $\phi$ (this is common to the $f(\mR)$ theories) and the Diff-breaking term expressed in the appearance of $\partial_\alpha q_{\mu\nu}$ that is new to the $f(\QQ)$ theories. It is worrisome that the Einstein--Hilbert term and the kinetic term for the conformal mode enter with the same sign, which can be seen as an indication of an unavoidable ghost either in the graviton or in the conformal mode. This is in high contrast with the metric $f(\mR)$ theories where the conformal mode enters with the correct sign. In~the $f(\QQ)$ theories, it is precisely the current $J^\mu$ that gives an extra contribution that flips the sign of the kinetic term $(\partial\phi)^2$. This does not preclude by itself a healthy conformal mode since the mixing with $q_{\mu\nu}$ in the last term could allow for a positive definite kinetic term. Thus, we can conclude that if the theory is to stay healthy, the~mixing between the conformal mode and the metric $q_{\mu\nu}$ from the diffeomorphism-breaking sector is crucial. By~applying the Sylvester criterion to the kinetic matrix, we conclude that it can only be positive definite if there are constraints that eventually flip the sign of $(\partial\phi)^2$ back to the healthy case.\footnote{Let us recall that a matrix is positive definitive if all the principal minors are strictly positive. Thus, avoiding ghosts forces all the diagonal elements of the kinetic matrix to be positive since the upper left element can be arbitrarily chosen.} On the other hand, we must notice the presence of both self-interactions and interactions with the conformal mode for the metrics that do not respect diffeomorphisms. This is generically a very pathological feature prone to containing a Boulware--Deser ghost~\cite{Boulware:1973my}. 

In view of the discussed drawbacks, it is interesting to note that the lapse and the shift in the ADM decomposition of $q_{\mu\nu}$ have vanishing conjugate momenta, i.e.,~they do not propagate at fully non-linear order. To~show this property, let us introduce the usual ADM decomposition for the metric
\be
q_{00}=-N^2+\gamma^{ij}N_iN_j,\quad q_{0i}=N_i,\quad q_{ij}=\gamma_{ij},
\ee
and its inverse
\be
q^{00}=-\frac{1}{N^2},\quad q^{0i}=\frac{N^i}{N^2},\quad q^{ij}=\gamma^{ij}-\frac{N^iN^j}{N^2},
\ee
with $\gamma^{ij}$ the inverse of $\gamma_{ij}$. The~determinant is given by $\det q_{\mu\nu}=-N^2\gamma$, with~$\gamma=\det \gamma_{ij}$. With~this decomposition, it is straightforward to see explicitly the non-dynamical nature of the lapse $N$ and the shift $N_i$ because the diffeomorphism-breaking sector is the only one that could potentially generate non-trivial momenta for them. Since the shift and the lapse mix with the derivatives of the conformal mode, there is a hope that integrating them out might flip the sign of the kinetic term. If~we consider a homogeneous configuration, that is the relevant case for our purpose, we can see that this sector gives the following contribution to the Lagrangian:
\be
\Lag\supset-\mpl\sqrt{-q}\left(q^{00}q^{ij}-q^{0i}q^{0j}\right)\dot{\phi}\,\dot{q}_{ij}=\mpl\sqrt{\gamma}\frac{1}{N} \gamma^{ij} \dot{\phi}\,\dot{\gamma}_{ij}
\ee
so neither the lapse nor the shift enter with time derivatives, thus guaranteeing a set of four primary constraints. This was already hinted in~\cite{Jimenez:2019ovq} in the original $f(\QQ)$ frame. In~the Einstein frame, the~non-dynamical nature of the lapse and the shift is slightly more transparent. Furthermore, we can use the identity $\partial_0\log\gamma=\gamma^{ij}\partial_0{\gamma}_{ij}$ to write the above expression as
\be
\Lag\supset-\mpl\sqrt{-q}\left(q^{00}q^{ij}-q^{0i}q^{0j}\right)\dot{\phi}\,\dot{q}_{ij}=\frac{\mpl}{\sqrt{\gamma} N} \dot{\phi}\,\dot{\gamma}.
\ee

It becomes apparent that actually only the determinant of the spatial metric mixes with the conformal mode. In \cite{Jimenez:2019ovq}, it was shown that the cosmological perturbations propagate two additional scalars with respect to GR, and below we will reproduce the same result from the perspective of the Einstein frame. In~view of the above partial results for the Hamiltonian analysis, it would seem reasonable to assign those two additional scalar modes to the conformal mode and the determinant of the metric in the Einstein frame. In~order to unveil the dynamical DOFs and, in~particular, the~presence of the Boulware--Deser ghost, a~Hamiltonian analysis would be necessary. The~Einstein frame \eqref{SCG} seems especially well-suited for this task that is beyond the scope of this communication, but~is currently under way~\cite{f(Q)Hamiltonian}.

In the matter sector, we will have the coupling to the usual conformal metric so the matter action in this frame is simply $\mS_{\rm m}=\mS_{\rm m}[\psi, e^{-2\phi/\mpl}q_{\mu\nu}]$. Among~other consequences, this conformal coupling could allow for a chameleon type of screening mechanism for the scalar field $\phi$ very much like in the $f(\mR)$ theories~\cite{Khoury:2003aq,Cembranos:2005fi}. The~impact of the Diff-breaking term in the scalar field sector should be analysed however before concluding the effective presence of this screening mechanism. In~particular, as~explained above, this term might be crucial to have a healthy conformal~mode.

Although one might object that the violation of Diffeomorphisms is caused precisely by our choice of coordinates, we should remember that the connection actually plays the role of Diff-St\"uckelbergs, and this should now be apparent from the action \eqref{SCG}. In~fact, the~singular property of the STEGR is that even in this gauge, Diffeomorphisms invariance arises as a gauge symmetry of the theory. The~reason is that the Legendre transformation becomes singular in that case and the Diff-breaking term does not appear from the onset. Coming back to the present case, the~Diff-breaking sector is expected to give rise to the propagation of more degrees of freedom. In~this frame, it is, however, straightforward to see that only the conformal mode will have a proper kinetic term around Minkowski. If~we take $\phi=\phi_0+\pi$ and $q_{\mu\nu}=\eta_{\mu\nu}+\frac{1}{\mpl}h_{\mu\nu}$. The~quadratic action reads
\be
\mS^{(2)}_{\rm CG}=\int\dd^4x\left[\Lag_{\rm GR}(h)+3(\partial\pi)^2-\frac12 m_\pi^2\pi^2-\Big(\partial^\alpha h-\partial_\beta h^{\alpha\beta}\Big)\partial_\alpha\pi\right]
\label{SCGquad}
\ee
with 
\be
\Lag_{\rm GR}(h)=-\frac18 \partial_\mu h_{\alpha\beta} \partial^\mu h^{\alpha\beta}+\frac14 \partial_\mu h_{\alpha\beta}\partial^\beta h^{\alpha\mu}-\frac14 \partial_\mu h\partial_\alpha h^{\alpha\mu}+\frac18 \partial_\mu h\partial^\mu h
\ee
the standard quadratic Lagrangian of GR, $h\equiv h^\alpha{}_\alpha$
and $m_\pi^2=\frac12\mpl^2\mU''(\phi_0)$ the mass of the field generated by the potential. We can now identify that a gauge symmetry appears on this background. If~we perform a Diff gauge transformation parametrised by $\zeta^\mu$ so that $\delta\pi=0$ and $\delta h_{\mu\nu}=2\partial_{(\mu}\zeta_{\nu)}$, we can see that it is a symmetry of the quadratic action. The~first three pieces in \eqref{SCGquad} are obviously invariant because they are Diff-invariant at the full-non linear level, so we have
\be
\delta_\zeta\mS^{(2)}_{\rm CG}= -\int\dd^4x\Big(\partial^\alpha \partial\cdot\zeta-\Box\zeta^\alpha\Big)\partial_\alpha\pi
\label{SCGquadzeta}
\ee
that vanishes upon integration by parts. The~symmetry becomes apparent by noticing that we can express the mixing between the conformal mode and $h_{\mu\nu}$ as $ h_{\mu\nu}J^{\mu\nu}$ with $J^{\mu\nu}=\partial^\mu\partial^\nu\pi-\eta^{\mu\nu}\Box\pi$ and identically off-shell conserved current.\footnote{This is the same kinetic mixing that appears between the St\"uckelberg field and the graviton in the decoupling limit of massive gravity~\cite{ArkaniHamed:2002sp}.} Since this symmetry is not present in the full theory, it is an accidental gauge symmetry of this background what signals that some propagating dofs become non-dynamical on this background and, consequently, this background suffers from strong coupling. In~other words, the~Minkowski solution will represent a singular surface in phase space that is likely to act as a repeller or singular (non-complete) trajectories in phase space so it will never be smoothly reached from an arbitrary point in phase space. Another interpretation of this feature is that the Cauchy problem will not be well-posed on this surface. This is consistent with the findings in~\cite{Jimenez:2019ovq}, where it was shown that some propagating modes in a general Friedmann--Lema\^itre--Robertson--Walker metric become non-dynamical around maximally symmetric backgrounds, in~particular Minkowski. We will obtain another version of this problem in a cosmological context~below.

The quadratic action can be diagonalised by performing the field redefinition
$h_{\mu\nu}\rightarrow h_{\mu\nu}+2\pi\eta_{\mu\nu}$, that corresponds to the linear version of a Weyl rescaling. After~this field redefinition, we obtain
\be
\mS^{(2)}_{\rm CG}=\int\dd^4x\left[\Lag_{\rm GR}(h)-\frac12 m_\pi^2\pi^2\right]
\label{SCGquaddiag}
\ee
so the kinetic mixing of the conformal mode with the graviton can be eliminated with the performed field redefinition without generating any kinetic term for $\pi$. In~this diagonalised action, it is trivial to see that we have a linearised Diffs gauge symmetry and the conformal mode is~non-dynamical.

It is instructive to reproduce some results of the cosmological solutions in this frame. The~background action in the mini-super space is given by
\be
\mS=\int\dd^3x\dd t a^3 N\left[\frac{3H^2\mpl^2}{N^2}\left(2+\frac{\dot{H}}{H^2}-\frac{\dot{N}}{H N}\right)-\frac{3}{N^2}\dot{\phib}^2-\UE(\phib) +6\frac{H\mpl}{N^2}\dot{\phib}\right].
\ee

The last term proportional to $\mpl$ is generated by the Diff-breaking term of the action and we see that it retains a time-reparametrisation invariance $t\rightarrow \alpha t$ together with $N\rightarrow N/\alpha$ that guarantees the existence of some Bianchi identities for the background cosmological equations. This is the Einstein frame version of the property already noted in~\cite{Jimenez:2019ovq}. Let us note that the relative coefficients between the two terms in (\ref{SCG}) is crucial for the existence of this symmetry, and this roots on the precise form of the current $J^\mu$. Any other relative coefficient between the coefficients of this current will break this symmetry, what shows the peculiarity of the non-metricity~scalar.

It is also instructive to recover the residual gauge symmetry found in~\cite{Jimenez:2019ovq} around maximally symmetric backgrounds in this frame. For~that, we will consider our gravitational action in the presence of a matter sector given by a canonical scalar field
\be
\mS_{\psi}=\int\dd^4x\sqrt{-q}\left[-\frac12e^{-2\phi/\mpl}q^{\mu\nu}\partial_\mu\psi\partial_\nu\psi-e^{-4\phi/\mpl}V(\psi)\right]
\ee
that we have already expressed in the conformally transformed frame, i.e.,~where the conformal mode couples directly to matter. We will now consider a perturbed cosmology described by the line element:
\be
\dd s^2_q=a^2\left[-(1+2\Phi)\dd \eta^2+2\partial_i B \dd x^i\dd \eta+\left(\big(1-2\Psi\big)\delta_{ij}+2\big(\partial_i\partial_j-\frac13 \delta_{ij}\nabla^2\big)E\right)\dd x^i\dd x^j\right]
\ee
with the conformal mode and the matter sector given by
\be
\phi=\phib+\delta\phi,\quad\quad\psi=\psib +\delta\psi.
\ee

We can obtain the equations of motion for the different perturbations and apply a gauge transformation of the form
\begin{eqnarray}
\Delta_\zeta\Phi&=&-(\zeta^0)'-\mH\zeta^0,\\
\Delta_\zeta\Psi&=&-\frac13k^2\zeta+\mH\zeta^0,\\
\Delta_\zeta B&=& -\zeta'+\zeta_0,\\
\Delta_\zeta E&=&-\zeta,\\
\Delta_\zeta\delta\phi&=&-\phib'\zeta^0,\\
\Delta_\zeta\delta\psi&=&-\psib'\zeta^0.
\end{eqnarray}

Upon use of the background equations of motion, the~corresponding perturbed equations change as
\begin{eqnarray}
\Delta_\zeta\mathcal{E}_\Phi&=&\phib' F_\Phi(\zeta_0,\zeta_z,\mH,\psib,\psib'),\\
\Delta_\zeta\mathcal{E}_\Psi&=&\phib' F_\Psi(\zeta_0,\zeta_z,\mH,\psib,\psib'),\\
\Delta_\zeta \mathcal{E}_B&=& -k^2\mpl\phib'\Big((\zeta^0)'+k^2\zeta\Big),\\
\Delta_\zeta \mathcal{E}_E&=&\frac43 k^4\mpl\phib'\zeta_0,\\
\Delta_\zeta\mathcal{E}_{\delta\phi}&=&-2k^2\mpl\mH (\zeta^0+\zeta'),\\
\Delta_\zeta\mathcal{E}_{\delta\psi}&=&0.
\end{eqnarray}
where $F_{\Phi,\Psi}$ are certain functions of their arguments whose precise form is not relevant for our purpose. The~transformation of the equation for the matter scalar field vanishes identically because it is diffeomorphism-invariant. Regarding the gravitational sector, we can see how the Diff-breaking terms result in the non-invariance of the corresponding equations. However, if~the conformal mode is constant on the background so $\phib'=0$, we see that we recover a residual gauge symmetry, provided the gauge parameters satisfy $\zeta^0+\zeta'=0$. This is the same residual symmetry identified in~\cite{Jimenez:2019ovq} for maximally symmetric backgrounds in the $f(\QQ)$ frame.\footnote{It is intriguing the constraint for the gauge parameters that reminds to the definition of a Killing vector. We are thus tempted to attribute this residual gauge symmetry to the additional isometries of maximally symmetric backgrounds.} In this frame, we see that those solutions correspond to having a constant conformal mode. Furthermore, if~we also have $\mH=0$ as it corresponds to Minkowski, the~equations do not change, in~accordance with our result above that the Minkowski background exhibits a general linearised Diff symmetry with no constraints on the gauge parameters. This further supports that the linearised gauge symmetry of Minkowski is an accidental symmetry of that~background.

\section{Metric~Teleparallelisms}
\label{Sec:MetricT}

We will now turn to the substantially more studied class of $f(\TT)$ theories. We can notice that it is straightforward to apply the procedure used to construct the Einstein frame of $f(\QQ)$ to the $f(\TT)$ theories. The~only difference with $f(\QQ)$ arises from the different boundary term, i.e.,~the different current $J^\mu$ that now is given by
\be
J^\mu=2T^{\beta}{}_{\alpha\beta}g^{\alpha\mu}=2e^{2\phi}T^{\beta}{}_{\alpha\beta}q^{\alpha\mu}.
\ee

Since the metric is determined in terms of the inertial connection parametrised by $\Lambda^\alpha{}_\mu$ so that $g_{\mu\nu}=\Lambda^\alpha{}_\mu\Lambda^\beta{}_\nu\eta_{\alpha\beta}$, we need to perform the conformal transformation on the fundamental field as $\Lambda^\alpha{}_\mu=e^{-\phi}\LambdaE^\alpha{}_\mu$ and the torsion acquires an inhomogeneous piece
\begin{equation}
    T^\alpha{}_{\mu\beta}= \TE^\alpha{}_{\mu\beta}-2\partial_{[\mu}\phi\;\delta^\alpha_{\beta]}.
\end{equation}

The current can then be written as
\be
J^\mu=2 e^{2\phi}\left(\TE^\mu-3 \partial^\mu\phi\right).
\ee

Obtaining the transformed action for the $f(\TT)$ theories then amounts to plugging this current into \eqref{SCG0} that yields
\begin{eqnarray}
\mS&=&\frac12\mpl^2\int\dd^4x\sqrt{-q}\left[\mR(q)+6(\partial\phi)^2-\tilde{\mU}(\phi)-4q^{\alpha\beta}\partial_\alpha\phi\TE_\beta\right]\nonumber\\
&=&\frac12\mpl^2\int\dd^4x\sqrt{-q}\left[\mR(q)+6(\partial\phi)^2-\tilde{\mU}(\phi)-8q^{\alpha\beta}\partial_\alpha\phi(\LambdaE^{-1})^\mu{}_\rho \partial_{[\beta}\LambdaE^\rho{}_{\mu]}\right].
\label{Eq:f(T)Conf}
\end{eqnarray}

We obtain the same kinetic term for the conformal mode, again with the wrong sign, and~the inertial connection $\LambdaE$ enters in the term that explicitly breaks local Lorentz invariance. The~wrong sign for the conformal mode was also noted in~\cite{Wright:2016ayu} and will have all the associated pathologies discussed above. We can also see from the action \eqref{Eq:f(T)Conf} that the Minkowski solution exhibits an accidental local symmetry, this time corresponding to a local Lorentz invariance. If~we take a Minkowski background with $\phi=\phi_0+\pi$ and consider perturbations along an $SO(3,1)$ direction for the connection, no quadratic kinetic terms appear for any DOF associated to the connection. This is because, for these perturbations, the metric remains Minkowski at all orders, so $q_{\mu\nu}=\eta_{\mu\nu}$ at all orders. On~the other hand, the~inertial connection can be parametrised in terms of an element of the Lorentz Lie algebra $\omega^\alpha{}_\beta\in \so$ as $\LambdaE=\exp\omega$ so the quadratic action reads
\begin{eqnarray}
\mS=-4\mpl^2\int\dd^4x\partial^\alpha\pi\partial_{[\alpha}\omega^\mu{}_{\mu]}
=-2\mpl^2\int\dd^4x\left(\partial^\alpha\pi\partial_{\alpha}\omega^\mu{}_{\mu}+\partial^\mu\partial^\alpha\pi\omega_{\mu\alpha}\right)=0
\end{eqnarray}
that is trivial because the generators of the Lorentz group are traceless and $\omega_{\mu\nu}=-\omega_{\nu\mu}$ and, consequently, we conclude that this corresponds to a local symmetry of the quadratic action around~Minkowski.

The theory preserves Diffs-invariance so there is a gauge symmetry at the full non-linear level. If~we parametrise a gauge transformation by $\zeta^\mu$, for~the considered Minkowski background now with $\LambdaE^\alpha{}_\mu=\delta^\alpha{}_\mu+\lambda^\alpha{}_\mu$, we have
\be
\delta_\zeta\pi=0,\quad\quad\delta_\zeta \lambda^\alpha{}_\mu=\partial_\mu\zeta^\alpha.
\label{eq:GaugTransflambda}
\ee

For these perturbations the action reads
\be
\mS^{(2)}=\int\dd^4x\left[\Lag_{\rm GR}(h)+3(\partial\pi)^2-\frac12 m_\pi^2\pi^2-2\Big(\partial_\alpha \lambda^\mu{}_\mu-\partial_\mu \lambda^\mu{}_\alpha\Big)\partial^\alpha\pi\right].
\label{SCGfTquad}
\ee
where we have rescaled $\pi\rightarrow\pi/\mpl$, $\lambda\rightarrow\lambda/\mpl$. It is then easy to see that the action is invariant under \eqref{eq:GaugTransflambda}.

We can split the connection perturbation into symmetric and antisymmetric pieces as $2\lambda_{\mu\nu}=h_{\mu\nu}+b_{\mu\nu}$. This decomposition isolates the pure metric perturbation $h_{\mu\nu}=2\lambda_{(\mu\nu)}$ from the perturbation $b_{\mu\nu}=2\lambda_{[\mu\nu]}$ along the $SO(3,1)$ direction at linear order. As~we have shown, the~quadratic action retains a local Lorentz symmetry, so $b_{\mu\nu}$ will not appear at this order, and the quadratic action can be written as
\be
\mS^{(2)}=\int\dd^4x\left[\Lag_{\rm GR}(h)+3(\partial\pi)^2-\frac12 m_\pi^2\pi^2-\Big(\partial^\alpha h-\partial_\beta h^{\alpha\beta}\Big)\partial_\alpha\pi\right].
\ee

We thus obtain the same quadratic action as in the Minkowski background for $f(\QQ)$ theories, and we need to go to higher orders to see the differences. As~shown above, the~kinetic mixing of the conformal mode can be absorbed into a field redefinition of $h_{\mu\nu}$, so we recover the standard result that no additional modes propagate around Minkowski. This property of the Minkowski background in $f(\TT)$ theories was already noticed in~\cite{Li:2011rn}, and the associated pathologies in relation with a problematic time evolution or ill-posedness of the Cauchy problem has been discussed in~\cite{Ong:2013qja,Izumi:2013dca}. A~detailed discussion on how background solutions with remnant symmetries (to use the language of the $f(\TT)$ literature) jeopardise the stability of those solutions can be found in~\cite{Chen:2014qtl}. Needless to say that all those shortcomings will also apply to the Minkowski solutions in the general class of teleparallel theories explored in this work, in~particular the $f(\QQ)$ theories studied above and the more general class that is the subject of the next~section.

\section{General~Teleparallelisms}
\label{sec:general}

Finally, it is straightforward to extend the above construction to the case of the general teleparallel theory. Let us consider the non-linear extension
\be
\mS=\frac12 \mpl^2\int\dd^4x\sqrt{-g} f(\GTGR).
\ee

In this case, the~total derivative involves both non-metricity and torsion, but~we can still apply the same procedure with the current
\be
J^\mu=Q^\mu-\Qb^\mu+2T^\mu.
\ee

Now there is no constraint between the inertial connection and the metric, so we can assign an arbitrary weight to the transformation law of $\Lambda$ and the connection transforms as
\be
\Gamma^\alpha{}_{\mu\beta}=\tilde{\Gamma}^\alpha{}_{\mu\beta}+w_\Lambda\partial_\mu\phi\delta^\alpha_\beta
\ee
that corresponds to an integrable projective transformation.\footnote{In the classification of scale transformations in metric-affine geometry~\cite{Iosifidis:2018zwo}, $w_\Lambda=0$ is the conformal transformation and $w_\Lambda=1$ is the frame rescaling.} This results in a transformation of the non-metricity as follows:
\be
Q_{\mu\alpha\beta}=
\nabla_\mu g_{\alpha\beta}=
\tilde{\nabla}_\mu q_{\alpha\beta}-2w_\Lambda\partial_\mu\phi q_{\alpha\beta}
\ee

The transformed current is then
\be
J^{\alpha}=e^{2\phi}\left[\Big(q^{\alpha\beta}q^{\mu\nu}-q^{\alpha\mu}q^{\beta\nu}\Big)\tilde{\nabla}_\beta q_{\mu\nu}+2\tilde{T}^\mu-6\partial^\mu\phi\right].
\ee

Notice that this is independent of $w_\Lambda$, and this explains why we obtained the same kinetic term for the conformal mode in $f(\QQ)$ and $f(\TT)$. Equipped with the general form of the current, we can write the action as
\be
\mS=\frac12 \mpl^2\int\dd^4x\sqrt{-q}\left[\mR(q)+6(\partial\phi)^2-\tilde{\mU}(\phi)-2\Big[\Big(q^{\alpha\beta}q^{\mu\nu}-q^{\alpha\mu}q^{\beta\nu}\Big)\tilde{\nabla}_\beta q_{\mu\nu}+2\tilde{T}^\alpha\Big]\partial_\alpha\phi\right].
\label{eq:f(G)Einstein}
\ee
where we can clearly see how the inertial connection only enters in the last term generated by the current $J^\alpha$, and its explicit form is
\begin{align}
\label{eq:f(G)EinsteinLambda}
&\mS_\Lambda=-\mpl^2\int\dd^4x\sqrt{-q}\left[\Big(q^{\alpha\beta}q^{\mu\nu}-q^{\alpha\mu}q^{\beta\nu}\Big)\tilde{\nabla}_\beta q_{\mu\nu}+2\tilde{T}^\alpha\right]\partial_\alpha\phi\\
&=-\mpl^2\int\dd^4x\sqrt{-q}\left[\Big(q^{\alpha\beta}q^{\mu\nu}-q^{\alpha\mu}q^{\beta\nu}\Big)\partial_\beta q_{\mu\nu}+\Big(\delta^\alpha_\mu q^{\beta\nu}-\delta^\nu_\mu q^{\alpha\beta}\Big)(\Lambda^{-1})^\mu{}_\rho\partial_\nu\Lambda^\rho{}_\beta \right]\partial_\alpha\phi.\nonumber
\end{align}

This final form for the action permits to generalise to the entire family of $f(\GTGR)$ theories in a straightforward manner the result mentioned above for $f(\QQ)$ (and known in the literature for $f(\TT)$) that the solutions with a constant conformal mode are the same as those of GR. Since the current couples to the derivative of the conformal mode, this will have no effect for a constant $\phi$ (i.e., for~a Lorentz invariant configuration), and its only contribution will come from the potential evaluated at that value as a shift in the value of the cosmological constant. This general coincidence of the space of solutions of GR and the considered teleparallel theories for a constant conformal mode already hints to the possibility of a pathological nature of the Minkowski background (with a constant conformal mode) as we show~next.

In order to proceed with our study of the Minkowski spectrum, let us consider a background with $q_{\mu\nu}=\eta_{\mu\nu}+\frac{1}{\mpl}h_{\mu\nu}$, $\phi=\phi_0+\frac{1}{\mpl}\pi$ and $\Lambda^\alpha{}_\mu=\delta^\alpha{}_\mu+\frac{1}{\mpl}\lambda^\alpha{}_\mu$ so the quadratic action reads
\be
\mS^{(2)}=\int\dd^4x\left[\Lag_{\rm GR}(h)+3(\partial\pi)^2-\frac12 m_\pi^2\pi^2-\Big(\partial^\alpha h-\partial_\beta h^{\alpha\beta}\Big)\partial_\alpha\pi-2\partial_\beta \lambda^{[\alpha\beta]}\partial_\alpha\pi\right].
\ee

Once again, we recover the corresponding accidental gauge symmetries around Minkowski. The~inertial connection only enters through its component along the Lorentz group $\lambda^{[\alpha\beta]}$ which, furthermore, completely drops upon integration by parts. Thus, the~inertial connection completely disappears from the quadratic action and can therefore be regarded as a pure $\GL$ gauge mode. It is important to notice that this is not a symmetry of the full theory, but~only of the linear theory. On~the other hand, the~metric perturbation acquires once again the Diff-invariant coupling to the scalar field through an identically conserved current, i.e.,~the conformal mode couples to the linearised Ricci scalar of $q_{\mu\nu}$ at this order. Since, as~discussed above, this can be absorbed by a (linear) Weyl rescaling of the metric, we obtain the same result as for $f(\TT)$ and $f(\QQ)$ that all additional degrees of freedom vanish at linear order around Minkowski due to the appearance of accidental gauge symmetries. While diffeomorphisms were expected because the full theory is Diff-invariant, the~local Lorentz symmetry is an accidental local symmetry of the Minkowski background. The~same pathological behaviour is thus expected for the whole class of $f(\GG)$ theories.

We can now gain a deeper understanding and establish a general result for non-linear extensions of teleparallel equivalents of GR. The~most general teleparallel equivalent is the one constructed with $\GTGR$, where only the defining property of teleparallel geometries is imposed. Then, the~two paradigmatic examples given by the STEGR and the TEGR are just two partially gauge-fixed actions that live under the umbrella of the general teleparallel equivalent of GR.\footnote{More logically, one would refer to this as TEGR (if not GTEGR), and~to metric teleparallel case as MTEGR since the symmetric teleparallel case is referred to as STEGR. However, in~this paper, we rather followed the historical convention than insisted on a new, logical convention.} In~this respect, it is then natural to expect that non-linear extensions based on these partially gauge-fixed versions of the theory share some common features. In~fact, the~partial gauge fixing can be imposed by introducing an appropriate gauge-fixing term in the action via, e.g.,~Lagrange multipliers~\cite{Hohmann:2021fpr}, and we can do this either in the original frame or in the Einstein frame. For~instance, the~Einstein frame of $f(\QQ)$ can be obtained from \eqref{eq:f(G)Einstein} or \eqref{eq:f(G)EinsteinLambda} by simply adding a Lagrange multiplier imposing $T^\alpha{}_{\mu\nu}=0$. Similarly, the~$f(\TT)$ theories can be obtained by imposing $Q_{\alpha\mu\nu}=0$ in \eqref{eq:f(G)Einstein} or \eqref{eq:f(G)EinsteinLambda}. However, since we have obtained the disappearance of any additional DOFs around Minkowski for the general $f(\GTGR)$, this result will remain for any other partially gauge fixed theory giving rise to genuinely distinctive non-linear~extensions.

\section{On the Existence of Inequivalent Minkowski~Solutions}
\label{sec:Minkowski}

In the preceding sections, we have provided a unified framework to understand the absence of additional DOFs on a Minkowski background for the considered non-linear extensions of the teleparallel equivalents of GR as a consequence of having accidental gauge symmetries. However, given the geometrical framework where these theories are formulated and, in~particular, the~loss of crucial symmetries in the non-linear extensions, it is pertinent to discuss how to properly characterise what we mean by Minkowski background and, in~this respect, how natural our {\it Minkowski Ans\"atze} are. We will focus on the symmetric and the metric teleparallelisms because they are the most extensively studied~cases. 

One could argue that there is a fundamental difference between the metric and the symmetric teleparallel theories owed to the different constraints imposed upon the connection. In~both teleparallelisms, the~connection is enforced to be flat so that it must be purely inertial, and it is thus determined by $\Lambda^\alpha{}_\mu$. In~the metric teleparallelism, the~additional metric compatibility constraint determines the spacetime metric in terms of this inertial connection. By~selecting a locally Lorentzian metric $\eta_{\alpha\beta}$, the~spacetime metric is then given~by
\be
g_{\mu\nu}=\eta_{\alpha\beta}\Lambda^\alpha{}_\mu\Lambda^\beta{}_\nu
\ee
i.e., the~inertial connection is nothing but the usual soldering form. This solution of the metric compatibility constraint leaves a local Lorentz symmetry in the determination of the metric from the inertial connection, which is at the origin of the possibility of constructing inequivalent solutions for $g_{\mu\nu}$. In~particular, it is evident that seeking for Minkowski solutions with $g_{\mu\nu}=\eta_{\mu\nu}$ only restricts the space of allowed inertial connections $\Lambda\in\GL$ to its Lorentz subgroup $SO(3,1)$. Thus, naively imposing a Minkowski metric does not completely determine the solution and different Lorentz-related inertial connections giving $g_{\mu\nu}=\eta_{\mu\nu}$ could lead to different linear spectra. This amounts to replacing the identity by an arbitrary element of $SO(3,1)$ in the expansion employed for the analysis of Section~\ref{Sec:MetricT}.

The symmetric teleparallelism exhibits an arguably different behaviour. In~this case, the~additional constraint imposes the connection to be torsion-free, and this leads to an inertial connection which further corresponds to a pure coordinate transformation. Thus, the~connection can be fully trivialised, the~metric tensor remains as the only dynamical variable, and there is no ambiguity in the definition of a Minkowski solution defined as a solution satisfying $Q_{\alpha\mu\nu}=0$. Since this is a tensorial identity, it does not depend on the specific gauge choice. In~the coincident gauge, it reduces to imposing the constancy of the metric coefficients. There is an interesting issue, however, as to how we are to define the Minkowski space regarding its connectivity. Everyone would agree that Minkowski is described by a metric whose associated Levi-Civita connection has a trivial Riemann tensor. However, this leaves the possibility of introducing any other connection that does not spoil this property. It seems natural to further stipulate that Minkowski corresponds to selecting a trivial connection. However, within~the symmetric teleparallel framework we can leave the inertial connection free (i.e., the~$\xi's$) while fixing the metric to be Minkowski, thus relaxing the stronger condition $Q_{\alpha\mu\nu}=0$. In~other words, we can look for non-trivial solutions for the Diff-St\"uckelberg fields imposing a Minkowski solution for the metric. Of~course, this would be equivalent to some non-trivial spacetime metrics by going to the coincident gauge, something that we can always do. The~question is somewhat analogous to the choice of inertial coordinates in special relativity. In~the symmetric teleparallel case, the same question arises, and a possible answer to this question has been given in~\cite{Jimenez:2019yyx}, where a prescription for a canonical frame is~given.

Clearly, we could {\it define} the Minkowski spacetime by the fully covariant property that the curvature, the~torsion and the non-metricity vanish. 
In the metric teleparallel theories, this amounts to the condition $T^\alpha{}_{\mu\nu}=0$ in analogy to the condition $Q_{\alpha\mu\nu}=0$ in symmetric teleparallel~theories.

Only this definition, coupled with the requirement that $g_{\mu\nu}=\eta_{\mu\nu}$, removes the ambiguity of the vacuum. It was first pointed out in the context of metric teleparallelism, that due to the breaking of the Lorentz invariance induced by the modification of gravity, the~vacuum as defined by $g_{\mu\nu}=\eta_{\mu\nu}$ is not unique, but~different choices of vacua, related by a Lorentz transformation, can have different particle spectra~\cite{Koivisto:2018loq}. This
was confirmed explicitly in the study of linear cosmological perturbations
in both flat and in spatially curved Friedmann--Lema\^{i}tre backgrounds
~\cite{Golovnev:2018wbh}. Most recently, Golovnev and Guzm\'{a}n presented a more
detailed case study in $g_{\mu\nu}=\eta_{\mu\nu}$ spacetime(s) using rotated and boosted tetrads, which resulted in the appearance of additional propagating mode(s)~\cite{Golovnev:2020nln}. However, the~condition $T^\alpha{}_{\mu\nu}=0$ proposed above eliminates the possibility of the non-trivial\footnote{By the trivial remaining freedom we more precisely mean a global $SO(3,1)$ in the metric teleparallelism, and~a global $GL(4,\mathbb{R})$ in the general teleparallelism. In~symmetric teleparallelism, there remains the global affine transformation of the coordinate system.} rotations and boosts. In~the thus defined vacuum, characterised by diagonal Minkowski tetrads,
the new modes are strongly coupled and manifest only at the 4th
order perturbation theory~\cite{Jimenez:2020ofm}.

\section{Discussion}
\label{sec:discussion}
In this paper, we have constructed appropriate frames for teleparallel theories based on non-linear extensions of the teleparallel scalars that provide alternative descriptions of GR in flat geometries. We have started with the most extensively studied cases of $f(\QQ)$ and $f(\TT)$ theories where the fundamental geometrical objects are the non-metricity and the torsion, respectively, but~we have also examined the $f(\GTGR)$ theories that contain both torsion and non-metricity. In~all cases, we have written the theories with the usual Einstein--Hilbert term, the~separated conformal mode (common to the metric $f(\mR)$ theories) and the sector that explicitly breaks the symmetries of the teleparallel equivalents of~GR. 

We have found that the conformal mode enters with the wrong kinetic term and couples derivatively to the symmetry-breaking sector. Thus, if~this mode is to be stable, a~crucial non-trivial mixing with that sector is needed. However, precisely the symmetry breaking nature of this sector makes it prone to pathologies, mainly in the form of ghostly DOFs like the Boulware--Deser ghost expected in $f(\QQ)$, although~this remains to be explicitly shown. We have instead focused on the linear spectra of the theories around Minkowski backgrounds. For~the $f(\QQ)$ theories, we have observed that the theory restores a diffeomorphism invariance that is not present in the full theory and, furthermore, the~conformal mode can be absorbed into a field redefinition so it becomes non-dynamical. Similarly, in~$f(\TT)$, we have observed that the quadratic action around Minkowski enjoys a linearised local Lorentz symmetry and, again, the~conformal mode becomes non-dynamical after an appropriate field redefinition. We thus recover that no additional modes besides the massless spin-2 propagates on Minkowski, and we have related this property to the existence of accidental gauge symmetries. It is important to emphasise that our results do not depend on the number and nature of the extra degrees of freedom in the full theory. Since the non-linear extensions break local symmetries, there will be associated extra modes and all of these will be strongly coupled on the Minkowski~background.

By considering the non-linear extensions based on the general teleparallel equivalent of GR, we have revealed how it is not a coincidence that $f(\QQ)$ and $f(\TT)$ share the same quadratic action around Minkowski since this is in turn what occurs for the whole class of theories based on non-linear extensions of teleparallel equivalents of GR. We have obtained the general quadratic action around Minkowski for the $f(\GTGR)$ theories and shown how they all exhibit sufficient accidental gauge symmetries as to guarantee that only the massless spin-2 mode propagates on Minkowski. Since the these symmetries are not present in the full theory, the presence of additional dynamical DOFs is expected, thus pointing towards the generic strongly coupled nature of Minkowski in these theories. The~existence of accidental gauge symmetries is shared by other theories formulated in the teleparallel framework, such as the class of general quadratic theories as discussed in, e.g.,~\cite{Hayashi:1979qx,PhysRevD.28.718,MullerNitsch85,Nester_1988,Blixt:2018znp}. In~fact, the~New General Relativity theory introduced in~\cite{Hayashi:1979qx} was defined precisely to have an accidental gauge symmetry on a Minkowski background (see the analysis in~\cite{ortin_2004}). The~accidental symmetry is the source of pathologies at the non-linear level as discussed in, e.g.,~\cite{Cheng:1988zg,Jimenez:2019tkx}. The~result for the non-linear extensions considered in our analysis is particularly remarkable because the found accidental symmetries ensure that no additional modes propagate on a Minkowski background, while other classes of teleparallel theories admit extra modes on Minkowski (a Kalb--Ramond field in the case of New GR for instance) even though it exhibits accidental~symmetries.

Finally, we have discussed the issue of equivalent physical solutions, especially the equivalence of Minkowski solutions within these theories highlighting the similarities, but~also the differences in metric and symmetric teleparallel theories. Conditions for the unambiguous definition of Minkowski spacetime are provided for both non-linear~extensions.

We will end this paper by noticing that our framework can be straightforwardly extended to scalar--teleparallel theories. For~instance, by~starting with the scalar-non-metricity action~\cite{Jarv:2018bgs,Runkla:2018xrv}
\be
\mS=\frac12\mpl\int\dd^4 x\sqrt{-g}\left[G(\psi) \QQ-(\partial\psi)^2-V(\psi)\right]
\ee
with $G(\psi)$ some function. It is apparent that we can rewrite this action in the form \eqref{eq:BransDickeF(Q)} via the field redefinition $\varphi=G(\psi)$ whenever $G$ is invertible. Then, we can directly go to the Einstein frame following exactly the same steps and the resulting action will be 
\begingroup\makeatletter\def\f@size{9}\check@mathfonts
\def\maketag@@@#1{\hbox{\m@th\normalsize\normalfont#1}}%

\be
\mS=\frac{\mpl^2}{2} \int\dd^4x\sqrt{-q}\left[\mR(q)+6\left(1-\frac{2e^{2\phi}}{3G'^2}\right)(\partial\phi)^2-\tilde{\mU}(\phi)-2\Big(q^{\alpha\beta}q^{\mu\nu}-q^{\alpha\mu}q^{\beta\nu}\Big)\partial_\alpha\phi\partial_\beta q_{\mu\nu}\right].
\label{Eq:EinsteinScalarNonmetricity}
\ee
\endgroup
Thus, we arrive at the same action but with a correction to the kinetic term of $\phi$. Obviously, the~remaining scalar-teleparallel extensions involving $\TT$ (see, e.g.,~\cite{Geng:2011aj,Jarv:2015odu,Hohmann:2018vle,Jarv:2021ehj}) and $\GTGR$ will exhibit similar properties. As~a matter of fact, it is evident that the Einstein frames will acquire exactly the same correction to the conformal mode as in \eqref{Eq:EinsteinScalarNonmetricity}, and the differences will arise again from the different currents $J^\mu$. The~correction to the kinetic term of the conformal mode can in turn be crucial for its stability since now we can always have a healthy region whenever $\frac{2e^{2\phi}}{3G'^2}>1$ that was not available in the non-linear extensions $F(\QQ)$, $F(\TT)$ and $F(\GTGR)$. Furthermore, the~extra contribution to the kinetic term of the conformal mode will also help it to propagate on a Minkowski background. This does not mean the complete absence of accidental gauge symmetries for these scalar--teleparallel extensions, but~shows again the maximal character of the non-linear theories in terms of symmetries or minimal in terms of propagating modes in the sense that only the graviton~propagates.

\acknowledgments{J.B.J.   acknowledges    support    from    the “Atracci\'on  del  Talento  Cient\'ifico”  en  Salamanca programme  and  from  project  PGC2018-096038-B-I00  by Spanish  Ministerio  de  Ciencia,  Innovaci\'on  y  Universidades. T.S.K. acknowledges support from the Estonian Research Council grants PRG356 “Gauge Gravity” and MOBTT86, and from the European Regional Development Fund CoE program TK133 “The Dark Side of the Universe”.}

\conflictsofinterest{The authors declare no conflict of interest.} 

\reftitle{References}


\bibliography{UniverseRefs}


%


\end{document}